\documentstyle[aas2pp4,iau200p]{article}

\begin{document}
\title{A Finding List for ${\bf\Delta\mu}$ Binaries \\
derived from a Comparison \\
of HIPPARCOS Proper Motions \\
with Long-term Averaged Data}
\author{Roland Wielen, Christian Dettbarn, Hartmut Jahrei{\ss},
Helmut Lenhardt, Heiner Schwan, Rainer J{\"a}hrling}
\affil{Astronomisches Rechen-Institut, \\
Moenchhofstrasse 12-14,
D-69120 Heidelberg, Germany\\
{\tt wielen@ari.uni-heidelberg.de}}

\begin{abstract}
The comparison of quasi-instantaneously measured HIPPARCOS proper motions with
long-term averaged proper motions, derived from ground-based data, allows the
identification of many stellar objects as $\Delta\mu$ binaries (Wielen et al.,
1999, A\&A 346, 675). We have used this method to find $\Delta\mu$ binaries
among the fundamental stars, among the GC stars, and among the Tycho-2 stars. A
finding list for about 4000 $\Delta\mu$ binaries is given under the URL
http://www.ari.uni-heidelberg.de/dmubin in a machine-readable format. This
information on the probable duplicity of the listed objects can be used for
planning specific observing programs, for improving our knowledge of binary
statistics, or for avoiding the use of these binaries in some astrophysical
calibrations
or as astrometric reference stars.
\end{abstract}

\section{Astrometric catalogues used}

In order to identify $\Delta\mu$ binaries (Wielen et al. 1999a) we compare the
HIPPARCOS proper motions (ESA 1997), which are quasi-instantaneously measured
within about three years, with long-term averaged, mean proper motions, which
are derived by using ground-based data from a much longer observational period
(often a century or more). If the difference $\Delta\mu$ between the short-term
and the long-term proper motion is statistically significant, the object is
probably a binary or a multiple system. The statistical significance of
$\Delta\mu$ is measured by the test parameter $F_{\Delta\mu}$. If the errors of
the proper-motion components were independent of the directions,
$F_{\Delta\mu}$ would be equal to $\Delta\mu$/(mean error of $\Delta\mu$). We
take into account the non-isotropy of the errors of $\Delta\mu$ and the
correlations between the components of $\Delta\mu$. A value of $F_{\Delta\mu}
>$ 3.44 corresponds to the same error probability as the familiar two-sided
3\,$\sigma$ criterion, and is therefore used by us for identifying $\Delta\mu$
binaries.

The long-term averaged proper motions are partially taken from available
astrometric catalogues; partially they are derived from the mean positions
given in the ground-based catalogues and the positions in the HIPPARCOS
Catalogue (we call this proper motion $\mu_0$). The ground-based catalogues are
corrected for systematic errors by comparing them with the HIPPARCOS Catalogue.

For determining various values of $\Delta\mu$ with respect to HIPPARCOS, we
have used the following astrometric catalogues: FK5 (Fricke et al. 1988, 1991),
RSup (Schwan et al. 1993), GC (Boss et al. 1937), and TYCHO-2 (Hoeg et al.
2000). More details will be given in other papers. For the basic FK5 stars,
Part I of the FK6 (Wielen et al. 1999b) gives already many additional
informations (see also URL: http://www.ari.uni-heidelberg.de/fk6).

\section{Finding list of ${\bf\Delta\mu}$ binaries}

A finding list for about 4000 $\Delta\mu$ binaries is given in the Internet
under URL\\[1ex]
\centerline{
{\bf http://www.ari.uni-heidelberg.de/dmubin}}\\[1ex]
in a machine-readable format. The list gives identification numbers, $F$
values, and other informations on these $\Delta\mu$ binaries. For many of these
objects, the probable duplicity was hitherto unknown. The highest fraction of
$\Delta\mu$ binaries occurs among the basic fundamental stars, because their
proper motions are most accurately measured. Among the fainter stars, the
fraction of identified $\Delta\mu$ binaries is smaller, due to the larger
measuring errors. The highest sensitivity of our $\Delta\mu$ method for
detecting binaries is reached for nearby stars.


\end{document}